\documentstyle[aps,prb,epsf,rotate,multicol]{revtex}

\def\va{{\bf a}}
\def\vb{{\bf b}}
\def\vc{{\bf c}}
\def\vd{{\bf d}}
\def\vp{{\bf p}}
\def\vq{{\bf q}}

\def\vA{{\bf A}}
\def\vH{{\bf H}}

\newcommand{\vsigma}{\mbox{\boldmath$\sigma$}}
\newcommand{\veta}{\mbox{\boldmath$\eta$}}
\newcommand{\vvarepsilon}{\mbox{\boldmath$\varepsilon$}}
\newcommand{\vDelta}{\mbox{\boldmath$\Delta$}}

\begin{document}

\title{Identification of the Orbital Pairing Symmetry in UPt$_3$}
\author{Matthias J.~Graf}
\address{Theory Division,
         Los Alamos National Laboratory,
         Los Alamos, New Mexico 87545}
\author{S.-K.~Yip$^{a,b}$ and J.~A.~Sauls$^a$}
\address{$^a$Department of Physics \& Astronomy,
         Northwestern University,
         Evanston, Illinois 60208
        \\
	$^b$Physics Division, 
	National Center for Theoretical Sciences, 
	Hsinchu 300, Taiwan}
\date{\today}
\maketitle
\draft
\begin{abstract}
This paper summarizes the results of a comprehensive analysis of the
thermodynamic and transport data for the superconducting phases of UPt$_3$.
Calculations of the transverse sound attenuation as a function of
temperature, frequency, polarization, and disorder are presented
for the leading models of the superconducting order parameter.
Measurements of the specific heat, thermal conductivity, and
transverse sound attenuation place strong constraints on the
orbital symmetry of the superconducting order parameter.
We show that the superconducting A and B phases are in excellent
agreement with pairing states belonging to the 
odd-parity E$_{2u}$ orbital representation.
\end{abstract}
\pacs{PACS numbers: 74.25.Ld, 74.25.Fy, 71.27.+a
\hfill LA-UR:99-4118}

\begin{multicols}{2}

\section{Introduction}

Unconventional superconductivity, the electronic analog of 
superfluidity in $^3$He, was discovered in the heavy-fermion metals
UBe$_{13}$ and UPt$_3$ more than a decade ago.\cite{ott83,ste84}
As in liquid $^3$He the observation of multiple superconducting
phases was the direct evidence for a multi-component superconducting order 
parameter.\cite{mul87,qia87,fis89,has89}
The phases of UPt$_3$ have since become a paradigm for
unconventional superconductivity. However, unlike the case of $^3$He the
identification of the orbital and spin symmetry of the order parameter 
has been a more difficult task. Heavy fermion metals are more complex
materials in which disorder, magnetism, spin-orbit coupling and anisotropy
must be factored into any realistic theory of superconductivity
in these systems (c.f. Refs.~\onlinecite{sau94,heffner96}).

In this paper we present new theoretical results and analysis of
the transport properties of the leading models for the superconducting phases
of UPt$_3$. These models yield qualitatively different predictions for the
transport properties in the superconducting phases.
We calculate the ultrasonic attenuation for the A and B phases
and discuss its sensitivity to order parameter symmetry, polarization direction
and disorder. From our analysis of experimental
data for the heat capacity,\cite{tai97} 
thermal conductivity\cite{lus96b,sud97}
and transverse sound attenuation\cite{ell96}
we determine the topology of the excitation gap on the Fermi surface
and conclude that the orbital symmetry of the order parameter in the A and
B phases of UPt$_3$ belongs to an odd-parity E$_{2u}$ representation.

\section{Pairing Symmetry}

The discoveries of multiple superconducting 
phases\cite{mul87,qia87,fis89,bru90,ade90}
of UPt$_3$ led to several
theoretical models for the superconducting phase diagram
based on different symmetry groups, or symmetry breaking
scenarios.\cite{hes89,mac89,sud89,luk91,che93}
One class of models is based on a two-dimensional 
(``E'') representation of the hexagonal
point group, $D_{6h}$, with the multi-component superconducting order 
parameter coupled to a {\sl symmetry breaking field} (SBF).
There are four E-representations for strong spin-orbit coupling: two E-reps
for both even-parity (E$_{1g}$, E$_{2g}$) and odd-parity (E$_{1u}$, E$_{2u}$)
pairing.  The E-rep models require a weak SBF
that lowers the symmetry of the normal state,
splits the superconducting transition and produces multiple
superconducting phases.\cite{hes89} The SBF is generally
assumed to be the in-plane antiferromagnetic order parameter that onsets at
$T_N\simeq 5\,K$;\cite{hay92} however, other explanations
of the SBF have been suggested.\cite{mid93,min93,ell97}
The precise structure of the short-range AFM correlations,
e.g. the spatial structure of domains, as well as the role of AFM as
a SBF for superconductivity is still an open question.\cite{mor00}
One of the outcomes of the calculations summarized below is that simple model
of equal-size, equally populated multi-domain structures for the SBF
is in disagreement with the anisotropy of the sound attenuation.

\begin{figure}
\noindent
\begin{minipage}{87mm}
\begin{center}
\mbox{
\epsfxsize=0.80\hsize
{\epsfbox{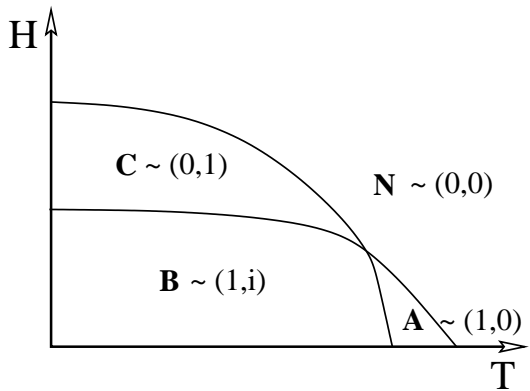}}
}
\vspace*{5mm}
\caption[]{The phase diagram of UPt$_3$. 
The three superconducting phases A, B, and C with amplitudes
$\veta = (\eta_1, \eta_2)$ meet with the normal state (N)
at the tetracritical point. 
For simplicity the additional two Meissner phases are not shown.
}
\label{phasediagram}
\end{center}
\end{minipage}
\end{figure}

The models that have been most successful in explaining the properties
of the superconducting phases of UPt$_3$
are based on the
even-parity (E$_{1g}$) and the odd-parity (E$_{2u}$) representations
of the hexagonal point group.
The E$_{1g}$ representation is a realization of
spin-singlet, d-wave pairing for a metal with a uniaxial symmetry,
while the E$_{2u}$ model describes the hexagonal
analog of spin-triplet, f-wave pairing. These pairing
states have an orbital order parameter of the form
$\Delta(\vp_f)=\eta_1 {\cal Y}_1(\vp_f)+\eta_2 {\cal Y}_2(\vp_f)$,
where ${\cal Y}_{1,2}(\vp_f)$ are the basis functions for the appropriate
E-representation, and the amplitudes $\veta=(\eta_1,\eta_2)$ transform as a
two-component `vector' under the same E-representation.
Thus, in the E-representations the order parameter of the A phase is
identified as $\veta=(1,0)$, the B phase as $\veta=(1,i)$, and the C phase as
$\veta=(0,1)$ (see Fig.~\ref{phasediagram}). These identifications
then refer to the specific basis functions, ${\cal Y}_{1,2}(\vp_f)$,
for a particular E-representation given in Table~\ref{OP_Reps}.
The orbital order parameter differs significantly for the two models,
particularly for the high temperature A phase $(\eta_2=0)$.
For E$_{1g}$ pairing the A phase
has the structure, $\Delta_{A}\sim p_z\,p_x$, which has an equatorial line node
in the basal plane, as well as a longitudinal line node circumscribing
the Fermi surface. For the E$_{2u}$ representation, $\Delta_{A}\sim
p_z\,(p_x^2-p_y^2)$ also has an equatorial line node, but has {\sl two}
longitudinal line nodes oriented $90$ degrees to one another. The
low-temperature B phase of both models breaks time-reversal symmetry
(with $\eta_2\simeq \pm i\eta_1$).
As a result the longitudinal line nodes are closed by the growth of the
second component of the order parameter; and for $T\rightarrow 0$,
$\Delta_B\sim p_z(p_x + i p_y)$ for E$_{1g}$ symmetry, while
$\Delta_B\sim p_z(p_x + i p_y)^2$ for the E$_{2u}$ representation.
Thus, the low energy excitation spectra for the B phase of the 
E$_{1g}$ and E$_{2u}$ models is
described by an equatorial line of zero energy excitations
($p_z=0$) and pairs of point nodes of the excitation
gap ($p_x=p_y=0$) on the Fermi surface. There is a slight difference
in the density of states from the point nodes because the gap
varies linearly near the point nodes for the E$_{1g}$ model,
$|\Delta(\vp_f)|\sim|\vartheta|$, but quadratically for
the  E$_{2u}$ model, $|\Delta(\vp_f)|\sim|\vartheta|^2$.
These slight differences are predicted to be observable
in the heat transport at ultra-low
temperatures.\cite{gra96b,nor96,fle95}

All of the E-rep models are based on two-component orbital order
parameters. However, they yield different predictions for the
thermodynamic, magnetic and transport properties,
including the H-T phase diagram.
One important difference arises for the case of weak 
in-plane hexagonal anisotropy, as is reflected by the
very small in-plane anisotropy of $H_{c2}$.\cite{shi86b,kel94}
Weak in-plane anisotropy leads to an apparent tetracritical
point for all field orientations provided the order parameter
belongs to an E$_{2}$ orbital representation.\cite{sau94}

A key difference between the even-parity and
odd-parity E-representations is the spin structure of the
order parameter. Odd-parity representations are spin-triplet
pairing states, and in the absence of spin-orbit coupling the
dimensionality of these representations is three times larger
than that of the corresponding spin-singlet E-representations.
However, strong spin-orbit coupling in the Uranium-based heavy fermion 
metals reduces the symmetry group by allowing 
only joint rotations of the spin and orbital degrees of
freedom. The even and odd parity representations are still described by
(pseudo) spin-singlet and spin-triplet 
order parameters of the form\cite{and84,vol85,blo85}
\begin{eqnarray}
\Delta_{\alpha\beta}({\vp}_f)=\Delta({\vp}_f)\,(i\sigma_y)_{\alpha\beta}
\qquad ({\rm singlet})\,,
\\
\Delta_{\alpha\beta}({\vp}_f)={\vDelta}({\vp}_f)\cdot
(i{\vsigma}\sigma_y)_{\alpha\beta}
\qquad ({\rm triplet})
\,.
\end{eqnarray}
The triplet representations transform only under {\sl joint}
spin and orbital rotations of the discrete point group for the normal 
state, i.e., $\vDelta(\vp_f)\rightarrow {\cal R} \vDelta({\cal R}^{-1}\vp_f)$,
where the rotation ${\cal R}\in[D_{6h}]_{\mbox{\tiny spin-orbit}}$.
The full symmetry group of the normal state is
${\cal G}=[D_{6h}]_{\mbox{\tiny spin-orbit}}\times{\cal T}\times U(1)$ 
with $[D_{6h}]_{\mbox{\tiny spin-orbit}}$ representing the hexagonal 
point group with inversion, ${\cal T}$ is the
time-inversion operation and $U(1)$ is the group of gauge transformations.
In the limit of no spin-orbit coupling $\vDelta(\vp_f)$ transforms
as a spin vector under the vector representation of the full 
spin-rotation group, and separately as a representation of the point
group with respect to the orbital momentum, $\vp_f$, i.e.,
$\vDelta(\vp_f)\rightarrow{\cal R}{\mbox{\tiny spin}}
\vDelta({\cal R}_{\mbox{\tiny orbit}}^{-1}\vp_f)$, 
where ${\cal R}_{\mbox{\tiny spin}}\in SU(2)_{\mbox{\tiny
spin}}$ and ${\cal R}_{\mbox{\tiny orbit}}\in 
[D_{6h}]_{\mbox{\tiny orbit}}$. In the absence of spin-orbit coupling
the enlarged symmetry group for the normal state is
${\cal G}= SU(2)_{\mbox{\tiny spin}}\times[D_{6h}]_{\mbox{\tiny
orbit}}\times{\cal T}\times U(1)$.

There are two special classes of spin-triplet order parameters that
are frequently discussed as candidates for the phases of UPt$_3$.
The first class are states in which the spin-triplet order parameter
factorizes into a single spin-vector and an orbital amplitude, i.e.,
$\vDelta(\vp_f)=\vd\,\Delta(\vp_f)$ where $\vd$ is a real unit vector
and $\Delta(\vp_f)$ is an odd-parity orbital function.
The vector $\vd$ defines the axis along which the 
pairs have zero spin projection, e.g., if $\vd || {\bf z}$ then
$\Delta_{\uparrow\uparrow}=\Delta_{\downarrow\downarrow}=0$ and
$\Delta_{\uparrow\downarrow}=\Delta_{\downarrow\uparrow}=\Delta(\vp)$.
If we choose the quantization axis to be perpendicular to $\vd$, i.e.,
$\vd\perp{\bf z}$, then
the same pairing state is described as {\sl equal spin pairing} in an
``easy-plane'', i.e., the pairs form triplet states with amplitudes
$\Delta_{\stackrel{\scriptstyle\leftarrow}{\scriptstyle\leftarrow}}=
 \Delta_{\stackrel{\scriptstyle\rightarrow}{\scriptstyle\rightarrow}}=
 \Delta(\vp)$
and $\Delta_{\stackrel{\scriptstyle\leftarrow}{\scriptstyle\rightarrow}}=0$. 
In the second class, $\vd$ is complex and the spin components of the 
order parameter spontaneously break time-reversal symmetry.
In the general case $\vDelta$ is complex, with
$\vDelta\times\vDelta^*\ne 0$, and
varies over the Fermi surface. These states are called `non-unitary'
because the square of the spin-matrix representation of the order parameter 
is no longer proportional to the unit spin matrix,
$[\Delta^{\dag}\Delta]_{\alpha\beta} = 
\left|\vDelta\right|^2\,\delta_{\alpha\beta} 
+i\left[ 
 \vDelta\times\vDelta^*\cdot\vsigma
 \right]_{\alpha\beta}$.
As a consequence the spin degeneracy of the excitation spectrum is lifted
and the quasiparticle energy depends on the local pair spin at $\vp_f$:
${\bf S}_{\mbox{\tiny pair}}(\vp_f)\sim i\vDelta(\vp_f)\times\vDelta(\vp_f)^*$.

Whether or not spin-orbit coupling is weak or strong on the energy
scale of $k_BT_c$ has important implications for both the
orbital and spin components of the order parameter that are allowed by
symmetry. Blount\cite{blo85} and Volovik and Gorkov\cite{vol85} 
showed that line nodes are {\sl not required}
for odd-parity states when spin-orbit coupling is relevant.
However, line nodes
in the $ab$-plane of the Fermi surface are allowed, and
required for some representations, \underline{if} the
normal-state spin-orbit interactions lock $\vd$
along the $\vc$ axis of the crystal, i.e., $\vd || \vc$.
Precisely this orientation of $\vd$ was predicted\cite{cho91} 
for UPt$_3$ based on anisotropic paramagnetic limiting.\cite{shi86b} 
This effect arises from the competition between the condensation
energy and the Zeeman energy. For $\vd$ locked along the $\vc$ axis
of the lattice the Zeeman energy is pair-breaking for $\vH || \vc$,
giving rise to paramagnetic limiting.
However, for $\vH\perp\vc$ 
the Zeeman energy, ${\cal F}_{\mbox{\tiny Zeeman}}\sim (\vd\cdot\vH)^2$,
is minimum (vanishes); as a result there is no paramagnetic limit 
for this field orientation. The anisotropic paramagnetic limiting
of $H_{c2}$ is sensitive to the spin structure of the order 
parameter, but insensitive to the orbital pairing 
symmetry.\cite{cho91,cho93,sau94,yang99}
The odd-parity E$_{2u}$ representation with strong spin-orbit locking
of $\vd ||\vc$ quantitatively accounts for the anisotropy of
the paramagnetic limit of $H_{c2}$ observed at low temperatures.

The spin-singlet E$_{1g}$ model appears to be incompatible with
both the tetracritical point for $\vH \not\perp \vc$ and the
anisotropic paramagnetic limiting of $H_{c2}$. However, Park and 
Joynt\cite{par96} argue that there is enough freedom in the E$_{1g}$
model to account for the existing experimental data on $H_{c2}$.
Both E-rep models have recently been challenged by
observations of a nearly temperature independent Knight shift for
$\vH || \vc$,\cite{tou98} which is interpreted in terms of non-unitary,
spin-triplet pairing with weak, or no spin-orbit coupling.\cite{mac98}
The authors of Ref.~\onlinecite{tou98} assume that the 
Knight shift measures the bulk spin susceptibility. If, for simplicity,
we ignore the anisotropy of the normal-state susceptibility, then
for a given orientation of $\vd$ the spin susceptibility is given by
$\chi_{ij}=\chi_N\,\left(\delta_{ij}-d_id_j\right)+\chi_{0}\,d_id_j$,
where $\chi_N$ is the normal state spin susceptibility and $\chi_0(T)$ is
the spin susceptibility for $\vH||\vd$, which is suppressed by pair-breaking
and vanishes for $T\rightarrow 0$ in the clean limit. 
For strong spin-orbit coupling with
$\vd$ locked along $\vc$ we expect a suppression of the Knight shift for
$T<T_c$ for fields $\vH || \vc$, but no suppression for $\vH\perp\vc$.
However, in the limit of no spin-orbit coupling the Zeeman
energy is minimized by {\sl rotation} of $\vd$ perpendicular to the
field. This implies that the Knight shift will be temperature independent
and given by the normal-state shift for all field orientations.
The NMR measurements of the Knight shift\cite{tou98} appear to be in conflict
with anisotropic paramagnetic limiting of $H_{c2}$. The paramagnetic
limit observed for $H_{c2}^{||}$ is a robust, thermodynamic property of bulk 
single crystals of UPt$_3$, and a consistent interpretation of the NMR
results for the Knight shift data must accommodate anisotropic 
paramagnetic limiting. 
This cannot be accomplished with a model of
spin-triplet pairing without strong spin-orbit coupling.

Our analysis presented below for the heat capacity and
transport measurements is independent of the interpretation
of the Knight shift measurements. We show that the heat capacity,
low temperature thermal conductivity and transverse sound
attenuation data, in addition to the H-T phase diagram, are in
quantitative agreement {\sl only} for the odd-parity E$_{2u}$ 
representation (Table \ref{OP_Reps}), 
independent of the orientation
of $\vd$. In order to demonstrate this fact we present calculations for other
models that have been proposed to account for the phase diagram.
Thus, in addition to the {\it 2D} E-representations we also examine
the transport properties of the order parameter models belonging
to mixed representations of the $D_{6h}$ point group, i.e.,
the AB models\cite{luk91,che93} and the AE model.\cite{zhi96}
These models were proposed as alternatives
to the E-representations to explain the Ginzburg-Landau region
of the H-T phase diagram. 
The most promising candidate of the AB model is
the odd parity, spin-triplet model with mixed A$_{2u}\oplus$B$_{1u}$
symmetry. The orbital order parameter for the A phase has the form,
$\Delta_A({\bf p}_f)\sim p_z {\rm Im}(p_x + i p_y)^6$, exhibiting
an equatorial line node and six longitudinal line nodes.
We also analyze the transport properties of the even-parity,
spin-singlet A$_{1g}\oplus$E$_{1g}$ model with an A phase of the form,
$\Delta_A({\bf p}_f) \sim (2p_z^2-p_x^2-p_y^2)$, which has a pair of
`tropical' line nodes located off the equatorial plane.
For a more detailed description of the order parameter for these
models see Refs.~\onlinecite{zhi96,gra99}.

\end{multicols}

%
%
\begin{center}
\begin{minipage}{\textwidth}
\begin{table}
\caption[]{
Polynomial functions representing the symmetry of the 
low-temperature B phases of several pairing models.
The first three entries are based on the symmetry group 
$[D_{6h}]_{\tiny spin-orbit}\times{\cal T}\times U(1)$.
The third entry is representative of the class of AB models,
and the last entry belongs to mixed symmetry representations
resulting from the crystal-field splitting of the enlarged symmetry 
group, $SO(3)_{\tiny spin-orbit} \times {\cal T} \times U(1)$.
}\label{OP_Reps}
\begin{tabular}{ccccc}
\noalign{\smallskip}
    {$\Gamma$}
  & {${\cal Y}_\Gamma$}
  & {point nodes}
  & {line nodes}
  & {cross nodes}
\\[1.0ex]
\noalign{\smallskip}\hline\noalign{\smallskip}
    $\rm E_{1g}$ & $p_z(p_x + i p_y)$
  & $\vartheta=0,\pi$   & $\vartheta=\frac{\pi}{2}$   & --
\\[1.0ex]
    $\rm E_{2u}$ & $ p_z (p_x + i p_y)^2$
  & $\vartheta=0,\pi$   & $\vartheta=\frac{\pi}{2}$    & --
\\[1.0ex]
    $\rm A_{2u} \oplus B_{1u}$
  & $\quad {\rm A}\,p_z{\rm Im\,} (p_x + i p_y)^6$      &
          --            & $\varphi_n=n\,\frac{\pi}{3}$,
                        & $\vartheta=0,\pi$ $\wedge\,\varphi_n$
\\
        & $\quad + {\it i}\,{\rm B}\,{\rm Im\,} (p_x + i p_y)^3 $
			&  
			&  $n=0,..,5$   
                        &  
\\[1.0ex]
    $\rm A_{1g} \oplus E_{1g}$
  & ${\rm A}\, (2 p_z^2 - p_x^2 - p_y^2)$ & -- & --
  & $\vartheta=\cos^{-1} \frac{\pm 1}{\sqrt{3}}$
\\
  & $+\quad {\it i}\,{\rm E} \, p_y p_z \quad$ & & &
    $\wedge\, \varphi=0, {\pi}$
\\
\end{tabular}
\end{table}
\end{minipage}
\end{center}

\begin{multicols}{2}

\section{Transport Theory}

Electronic transport in the superconducting state
is sensitive to the nodal structure of the order parameter,
$\Delta(\vp_f)$. Recent theoretical analyses\cite{gra96a,gra96b,nor96,gra99} of
low-temperature thermal conductivity data on superconducting
UPt$_3$\cite{lus96b,sud97} have eliminated most of the theoretical
models proposed to explain the phase diagram of UPt$_3$.
The non-unitary, spin-triplet pairing states
based on a one-dimensional ({\it 1D}) orbital representation
studied so far,\cite{mac91,ohm93,mac93,mac96,mac98} as well as
the two-component order parameter models
obtained from nearly degenerate one-dimensional representations\cite{che93}
(`AB models'), are unable to describe, even qualitatively, the temperature
dependence and anisotropy of the thermal conductivity at low temperatures.
The only pairing models which can account for the thermal conductivity data are
the two-dimensional ({\it 2D}) orbital representations,
E$_{1g}$ and E$_{2u}$, and the A$_{1g}\oplus$E$_{1g}$ (AE) model.
However, the AE model predicts a large $ab$-plane anisotropy, which 
has so far not been observed.\cite{gra99}
After it was shown that a non-unitary, spin-triplet state 
with a {\it 1D} orbital basis function was incompatible with
the thermal conductivity data,\cite{gra99} Machida et al.\cite{mac99} modified
their weak spin-orbit coupling model by adopting the {\it 2D} orbital
representation E$_{2u}$. However, the model of
Ref.~\onlinecite{mac99} proposes a spin structure for the order parameter
which is in conflict with the observed Pauli limiting for $\vH || \vc$, and
it predicts a fourth superconducting phase which disagrees with the phase
diagram.

Transverse ultrasound is an even more powerful probe of the order parameter and
excitation spectrum than the thermal
conductivity.\cite{shi86a,ell96} The attenuation of hydrodynamic sound
is determined by the electronic viscosity tensor, which is sensitive to the
relative orientation of the polarization, propagation direction and order
parameter.\cite{sch86,hir86}
The broken symmetries of the pairing state give rise to additional
anisotropy of the low-energy excitation spectrum that is specific to the pairing
state; the selection rules for acoustic absorption reflect these broken
symmetries.

In the hydrodynamic limit, $\omega\tau \ll 1$ and $q \ell \ll 1$, 
where $\ell=v_f\tau$ is the quasiparticle mean-free-path and $\tau$ is the 
transport collision time, the ultrasonic attenuation is determined by
components of the viscosity tensor
\begin{equation}\label{def_attenuation}
        \alpha(\vq,\vvarepsilon,T)
= ({\omega^2}/{\varrho c_s^3}) \, 
\eta_{i j, k l}({\bf q}, \omega)\,
	\hat{\vvarepsilon}_i \hat\vq_j
	\hat{\vvarepsilon}_k \hat\vq_l \,,
\end{equation}
where $\varrho$ is the mass density, $c_s=\omega/q$ is the 
speed of the sound mode with wavevector $\vq$ and polarization 
$\vvarepsilon$.\cite{mas55,lan59a}
The hydrodynamic limit is achieved even for high-purity single-crystals of
UPt$_3$. 
For the experiments reported in Ref.~\onlinecite{ell96} 
with propagation $\vq || \va$ and polarization $\vvarepsilon || \vb$
the sound frequency is
$\omega/2\pi\simeq 165\,\mbox{MHz}$, the speed of sound is
$c_s \simeq 2.1\, {\rm km/s}$,\cite{shi86b} 
and the elastic mean free path is 
$\ell_{ab}=v_{f,ab}\tau\approx 1.5\,\mbox{km/s}\cdot
240\,\mbox{ps}\approx 360\,\mbox{nm}$,\cite{kyc98}, yielding 
$\omega\tau\approx 0.25$ and $q\ell_{ab}\approx 0.18$ at $T=0$.
Similarly, for $\vq || \va$ and $\vvarepsilon || \vc$ the reported values are
$\omega/2\pi\simeq 228\,\mbox{MHz}$,
$c_s \simeq 1.4\, {\rm km/s}$, and $\ell_{c} \approx \sqrt{2.7} \ell_{ab}$,
yielding 
$\omega\tau\approx 0.34$ and $q\ell_{c}\approx 0.61$ at $T=0$.
The parameters $q\ell_c$ and $\omega\tau$ are a factor of two smaller
near $T_c$ than they are at low temperature, since $\tau(T_c)\simeq\tau(0)/2$.
Nevertheless, attenuation measurements for $\vvarepsilon || \vc$ are near
the borderline of the hydrodynamic regime. Measurements above and below
this cross-over regime would be desirable; both for checking the applicability
of hydrodynamic results for the attenuation for $\vvarepsilon || \vc$, and 
such measurements might exhibit to new phenomena in the collisionless regime.

The viscosity and sound attenuation 
are calculated from the response of the momentum stress tensor to an
ionic displacement field $\vA(\vq,\omega)=A(\vq,\omega)\vvarepsilon$.
For transverse modes ($\vq \cdot \vvarepsilon=0$) the
stress and viscosity tensors are related in the hydrodynamic 
limit by\cite{tsu60a,kad64}
\begin{equation}\label{def_viscosity}
\Pi_{i j}({\bf q},\omega)=\omega\,q\, A(\vq,\omega) \,
        \eta_{ij,kl}(\vq,\omega) \,
        \hat{\vvarepsilon}_k \hat{\vq}_l 
\,.
\end{equation}

At low temperatures the transfer of energy and momentum between the
ionic lattice and electronic excitations is dominated by the scattering
of quasiparticles off impurities or defects.
The theory of momentum transport by
quasiparticle scattering is formulated in terms of nonequilibrium 
Green's functions for electronic quasiparticles coupled to the acoustic 
modes of the lattice. The momentum stress tensor is
\begin{equation}\label{stress_tensor}
\hspace*{-1mm}\Pi_{ij}({\vq},\omega)= 
        N_f\!\int\!\!\frac{d \epsilon}{8\pi i}
	\int\!\!d\vp_f\,[{\bf v}_f]_i[\vp_f]_j
	\delta g^K\!(\vp_f,\vq; \epsilon,\omega),
\end{equation}
where ${\bf v}_f$ is the Fermi velocity, $\vp_f$ is the Fermi momentum, 
$N_f$ the density of states at the Fermi surface, and $\delta g^K$
is the nonequilibrium quasiparticle Green's function,
integrated with respect to the quasiparticle energy,
$\xi_{\vp}\simeq v_f(p-p_f)$; $\delta g^K$ includes both the changes
in the {\sl distribution} of occupied states and the dynamics
of the {\sl spectrum} of low-energy excitations (see Appendix).

\section{Transport Equations}

We use Keldysh's formulation of the nonequilibrium response
theory and calculate the transport properties and related Green's functions
in the quasiclassical limit, which is easily achieved in UPt$_3$ for excitation
energies, $(k_B T, \hbar\omega)\ll E_f\approx 1\,\mbox{meV}$, and wavelengths
long compared to the Fermi wavelength, i.e., $\hbar q\ll p_f$.
The central equation for the nonequilibrium Green's function
in the quasiclassical limit is a transport equation. For small
deviations from equilibrium the transport equation
may be linearized in the deviations of the Green's function from
its local equilibrium form. Our analysis and notation follows that of
Ref.~\onlinecite{gra96a}, which provides a detailed discussion
of the quasiclassical linear response theory, including a complete
solution to the {\sl linearized} nonequilibrium
transport equations. A summary of these equations, applicable to 
momentum transport in Fermi-liquid superconductors,
is given in the Appendix.

The transport equation
for the Green's function includes the acceleration of
electronic quasiparticles by the acoustic field and collision
terms which transfer momentum between the lattice and the electrons.
The stress tensor, and therefore the electronic viscosity which damps
the acoustic wave, is calculated from the solution of the transport equation
for the nonequilibrium Green's function, $\delta g^K$, which is driven by
the coupling of quasiparticles to the ionic displacement field,
i.e., an externally imposed sound field,
$\hat{\sigma}_{\mbox{\tiny ext}}(\vp_f,\vq;\omega)
=i({\bf v}_f\cdot\vq)(\vp_f\cdot\vA)\hat{1}$.\cite{tsu60a,kad64}
Below we report new results for the electronic shear viscosity for the
order parameter models of UPt$_3$, and new calculations of the anisotropy and
temperature dependence of the attenuation which we use to
interpret the experimental data for UPt$_3$.

In the limit of $\omega \to 0$ and for resonant scattering
the viscosity tensor simplifies to
\begin{eqnarray}\label{viscosity}
&&\eta_{ij,ij} = -\frac{N_f}{8 \pi^3 k_B T}
 \int d\epsilon\,{\rm sech}^2(\epsilon/2 k_B T)\,\times
\nonumber
\\
&&\int d\vp_f
\frac{[{\bf v}_f]_i^2 [\vp_f]_j^2}{{\rm Re\,}C^R}
\left[g_0^R g_0^{R*}-f_0^R f_0^{R*}+\pi^2\right]
\,,
\end{eqnarray}
where 
$C^R = -\frac{1}{\pi}\sqrt{|\Delta(\vp_f)|^2 - (\tilde\epsilon^R)^2}$,
$g_0^R = \tilde\epsilon^R/C^R$, $f_0^R=-\Delta(\vp_f)/C^R$, 
and $\tilde{\epsilon}^R= \epsilon-{\scriptsize 1\over 4}\mbox{Tr}
 [\hat{\tau}_3\hat{\sigma}^R_{\mbox{\tiny imp}}]$
is the impurity-renormalized energy. In the case of triplet pairing with
a unitary order parameter the only change in Eq.~(\ref{viscosity})
is replacement of $\Delta\rightarrow\vDelta$ and
$f^R_0f^{R*}_0\rightarrow {\bf f}^R_0\cdot {\bf f}^{R*}_0$.
See the Appendix for details on the notation.

For a normal metal with a spherical Fermi surface we obtain Pippard's
result for the viscosity,
$\eta_{i j,i j}={\scriptsize 2\over 15} v_f^2 p_f^2 N_f\tau$ for $i\ne j$.\cite{pip55}
Below $T_c$ the sound attenuation drops; for a conventional superconductor,
in the limit $q \ell \ll 1$, the attenuation decreases
exponentially for $k_BT\ll\Delta$.\cite{tsu60a}
But for an unconventional superconductor in which the
order parameter vanishes at points or lines on the Fermi
surface, the attenuation decreases with temperature as a power
law reflecting the spectrum of low-energy excitations near
the nodes of the order parameter.\cite{sch86,hir86,pet86,mon87a}
Impurity scattering modifies this spectrum near the nodes, and
at low energies a new energy scale, $\gamma\ll\Delta_0$,
appears, which is roughly the `bandwidth' of low-energy quasiparticles bound
to the impurity distribution by Andreev scattering.\cite{gra96a}
The bandwidth also appears as
an impurity-renormalized quasiparticle width at zero energy,
$\tilde\epsilon^R(0) = i\gamma$.
This new low energy scale defines a
cross-over from the power law behavior associated with scattering of continuum
quasiparticles for $\gamma/k_B<T\ll T_c$, 
to a temperature independent attenuation
in the limit $k_B T\ll\gamma$.\cite{sch86,hir86,pet86,mon87a}

The bandwidth of the impurity-induced Andreev levels is determined
by the self-consistency equation for the quasiparticle
self-energy,
\begin{equation}
\gamma=\Gamma_u\,
\frac{\langle\gamma\left[|\Delta(\vp_f)|^2+\gamma^2
\right]^{-{\scriptsize 1\over 2}}\rangle}
{\cot^2\delta_0 +
\langle\gamma\left[|\Delta(\vp_f)|^2+
\gamma^2\right]^{-{\scriptsize 1\over 2}}\rangle^2}
\,,
\end{equation}
where $\langle ...\rangle$ is an average over the Fermi surface
and $\Gamma_u=n_{\mbox{\tiny imp}}/\pi N_f$ is the scattering
rate in the normal state for resonant impurities and $\delta_0$ is 
the scattering phase shift.
For the high purity UPt$_3$ crystals studied in Ref.~\onlinecite{lus96b},
i.e., low scattering rate, the
crossover temperature is very low compared to $T_c$; analysis of the
thermal conductivity provides a determination of both the scattering
phase shift as well as the bandwidth of the impurity-induced Andreev states.
The scattering centers are nearly resonant, i.e., $\delta_0\simeq\pi/2$,
giving a bandwidth, and cross-over temperature, of order
$\gamma\sim k_B T^*\approx 0.2\sqrt{\mu\Gamma_u\Delta_0}\approx 0.07 k_B T_c$,
where $\Gamma_u\simeq 0.03 k_B T_c$, $\Delta_0\simeq 2.0 k_B T_c$ and
the slope of the excitation gap near the line node is
$\mu=\Delta_0^{-1} |d\Delta/d\vartheta|_{\vartheta=\pi/2}\simeq 2$.
Thus, transport experiments on UPt$_3$ have so far not
investigated the ultra-low temperature
region $k_B T \ll \gamma$ in any systematic way.

\section{Transverse Sound Attenuation}

\begin{figure}
\noindent
\begin{minipage}{87mm}
\begin{center}
\mbox{
\epsfysize=0.90\hsize \rotate[r]{\epsfbox{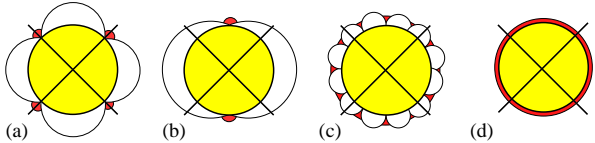}}
}
\vspace*{5mm}
\caption[]{Sketch of the $\phi$ dependence of the
A phase pairing states at $\vartheta\ne \pi/2$ 
for (a) E$_{2u}$, (b) E$_{1g}$, (c) 
$A_{2u}\oplus B_{1u}$, and (d) $A_{1g}\oplus E_{1g}$
models. The crossed lines represent the areas of
maximal absorption of a sound probe with $\va\vb$ 
symmetry ($\vq||\va, \vvarepsilon||\vb$)
at the Fermi surface. The dark shaded areas 
represent the distribution of quasiparticle excitations.
}
\label{NodeLocking-xy}
\end{center}
\end{minipage}
\end{figure}

The {\sl anisotropy} and temperature dependence of the sound attenuation 
is sensitive to the polarization of the sound field and the symmetry of
the order parameter. This was the basis
of transverse sound attenuation experiments that provided early evidence for
a line of nodal excitations in the basal plane.\cite{shi86a}
We examine the $ab$-plane anisotropy of the transverse
sound attenuation. Our analysis covers the full temperature range below $T_c$, 
and is particularly sensitive to the polarization and and anisotropy of the 
order parameter for both A and B phases of UPt$_3$.
To illustrate the sensitivity of the transverse sound polarization
to the order parameter symmetry consider the theoretical models
for the A phase of UPt$_3$. The $ab$-plane anisotropies
of the excitation gap, $|\Delta(\vartheta\ne\pi/2,\phi)|$, for the
A phase of four pairing models, E$_{2u}$, E$_{1g}$, 
$A_{2u}\oplus B_{1u}$ and $A_{1g}\oplus E_{1g}$, are shown
in Fig.~\ref{NodeLocking-xy}.

The propagation and polarization vectors determine the
angular dependence of the momentum transport by
quasiparticles on the Fermi surface; the matrix element
is proportional to $[\vp_f]_i^2[{\bf v}_f]_j^2$. This angular
dependence is weighted by the angle-resolved density of
states for momentum transfer via  impurity scattering,
which depends on the anisotropy of the quasiparticle
excitation spectrum through $|\Delta(\vp_f)|^2$.
When both the propagation and polarization vectors are
in the basal plane ($\vq||\va, \vvarepsilon||\vb$)\cite{coordinates}
the matrix element is proportional to 
$\sin^2 2\phi$, and is maximum at angles
of $\pi/4$ from these two axes, i.e., the midpoints between the
polarization and propagation directions.
If these midpoint directions coincide with nodal directions 
(e.g. $|\Delta(\vp_f)|\sim |\cos 2\phi|$ for E$_{2u}$)
then the attenuation
will be a maximum, while if the midpoint directions are along
the antinodal directions then the attenuation is a minimum.
This is illustrated in Fig.~\ref{NodeLocking-xy} where the
polarization is directed along the $\va$ direction. The attenuation 
is largest when the {\sl polarization} is along an antinode of
the order parameter, and it is smallest when the polarization
is along a nodal direction. One can immediately see that 
we should expect to observe a rather different $ab$-plane 
angular dependence to the attenuation for the different order
parameter models proposed for UPt$_3$. We quantify these
remarks below.

\subsection{Results}

In order to make quantitative predictions for UPt$_3$ we use
heat capacity and thermal conductivity
measurements to fix the magnitude of the order parameter, the 
Fermi surface anisotropy, the nodal parameters and the scattering
rate, all of which control the temperature dependence and anisotropy
of the thermal conductivity below $T_c$.\cite{gra96b,gra99}
We used variational basis functions based on
the polynomial functions in Table \ref{OP_Reps}. Symmetry dictates the
geometry of the line and point nodes, as well as the topological indices for
the point nodes, but not the slopes or curvature of the nodal regions of
the excitation gap. Our variational procedure fits these parameters
to the low-energy excitation spectrum from the thermal conductivity
data of Ref.~\onlinecite{lus96b}. The order parameter is then determined
self-consistently. For a detailed discussion of this analysis see 
Refs.~\onlinecite{gra96b,gra99}.
The anisotropy and temperature
dependence of the transverse sound attenuation are then calculated
with no additional parameters 
or adjustments of the order parameter models shown
in Fig. \ref{NodeLocking-xy}.

\begin{figure}
\noindent
\begin{minipage}{87mm}
\begin{center}
\mbox{\epsfxsize=0.85\hsize {\epsfbox{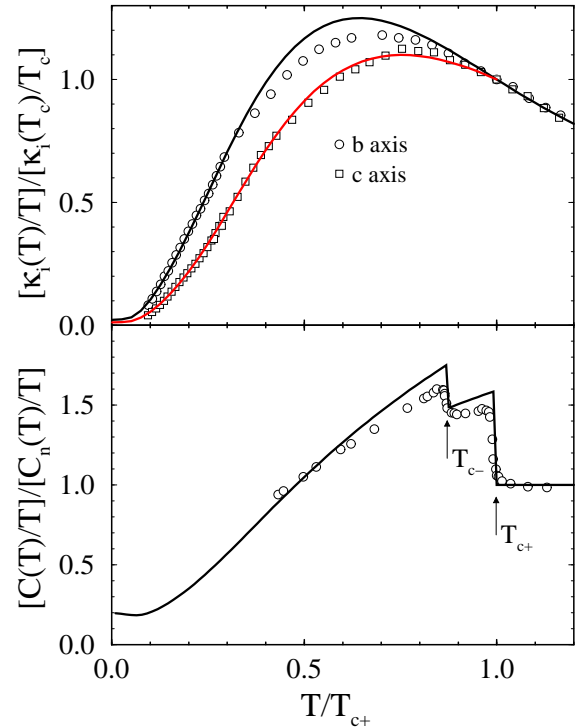}}}
\caption[]{Fit of the E$_{2u}$ model to the thermal conductivity 
 data (top)\cite{lus96b} and the specific heat data (bottom).\cite{tai97}
}\label{Fit}
\end{center}
\end{minipage}
\end{figure}

The transition temperature and splitting of the zero-field transition
in UPt$_3$ determine the scale and relative magnitudes of the two 
order parameter components; they determine the dominant 
and subdominant instability temperatures. The instability temperature 
for the dominant channel is the transition temperature, $T_{c1}=T_{c+}$,
while the second instability temperature
represents the strength of the subdominant pairing channel,
$T_{c2}\propto\omega_{c}\exp(-1/V_{2})$, where $V_{2}$ is the
pairing interaction for the sub-dominant channel
and $\omega_c$ is the cutoff energy.\cite{footnote2}
The physical transition temperature, $T_{c-}$, separating the 
A and B phases depends on the relative strength of the two paring
channels, i.e., on $T_{c2}/T_{c+} \le 1$.
We solve the gap equation for the second transition and
adjust $T_{c2}$ to the observed splitting in the specific heat,
$(T_{c+}-T_{c-})/T_{c+}\approx 60\,{\rm mK}/495\,{\rm mK}$.\cite{ell96,tai97}
All other material parameters are taken from our previous analysis
of the thermal conductivity of UPt$_3$.\cite{gra99} 
We show the quality of the theoretical fits to the thermal conductivity,
$\kappa$, and heat capacity for the E$_{2u}$ model in Fig.~\ref{Fit}. 
Similar fits can be obtained
for the E$_{1g}$ and AE models. However, only the in-plane thermal 
conductivity data can be accounted for by the AB model.

\subsubsection{Identification of the A-phase}

\begin{figure}
\noindent
\begin{minipage}{87mm}
\begin{center}
\mbox{\epsfysize=0.95\hsize\rotate[r]{\epsfbox{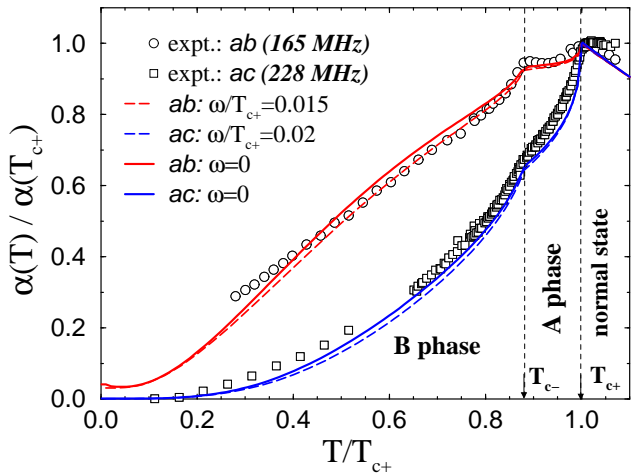}}}
\caption[]{Comparison between the measured and calculated transverse 
           sound attenuation of UPt$_3$. The theoretical calculation
	   is for the E$_{2u}$ model and the 
	   phenomenological model previously used for the
	   total elastic and inelastic
	   scattering rate, $\Gamma(T) = 0.03 k_B T_{c+}
	   (1+T^2/T_{c+}^2)$,
	   obtained from the thermal conductivity
	   analysis. The data are from Ellman {\it et al.},\cite{ell96}
	   which are corrected to vanish at $T=0$.
}\label{UPt3}
\end{center}
\end{minipage}
\end{figure}

In Fig.~\ref{UPt3} we show the attenuation data reported
by Ellman {\it et al.}\cite{ell96} for transverse sound propagation
in the $ab$-plane with $\vq||{\va}$ and polarizations
both in- and out-of-plane,
$\vvarepsilon||{\vb}$ and $\vvarepsilon||{\vc}$.
These measurements were made on the same batch of crystals of UPt$_3$
as the heat capacity and thermal conductivity measurements
shown in Fig.~2. In addition to the anisotropy
associated with the polarization, the data show a pronounced
change in the anisotropy and temperature dependence at the A-B
phase transition.

The enhanced absorption in the
A phase compared to that of the B phase at the same temperature
results from excess quasiparticles that scatter
off the impurity distribution due to additional nodes of the 
A phase order parameter compared to that of the B phase.
Below $T_{c-}$ the subdominant order parameter
closes the additional nodes of the A phase order parameter.
Thus, the sound absorption drops faster in the B phase than it would
in the A phase. The experimental results for the sound attenuation,
including the anomaly at $T_{c-}$, are in excellent agreement
with theoretical calculations for the E$_{2u}$ model of the
A and B phases, with an A phase given by a $\veta = (1,0)$ state,
but not a $(0,1)$ state; and a B phase at low temperature 
which is (approximately) the $\veta = (1,\pm i)$ state.\cite{gra00}
The comparison between theory and experiment
is unsatisfactory for all other models.
Even including the frequency dependence of the viscosity, 
$\eta_{ij, ij}(\vq,\omega)$, does not change this result.
We find only minor corrections to the sound attenuation 
for $\omega\ll \Delta_0$ (see Fig.~\ref{UPt3}).
The $(1,0)$ state of the A phase, as determined by
sound attenuation, also agrees with the order parameter orientation
obtained from the observed six-fold
oscillations of $H_{c2}(\phi)$, and the change in sign of these oscillations
when crossing the A-C phase boundary.\cite{sau96}

\begin{figure}
\noindent
\begin{minipage}{87mm}
\begin{center}
\mbox{\epsfysize=0.97\hsize\rotate[r]{\epsfbox{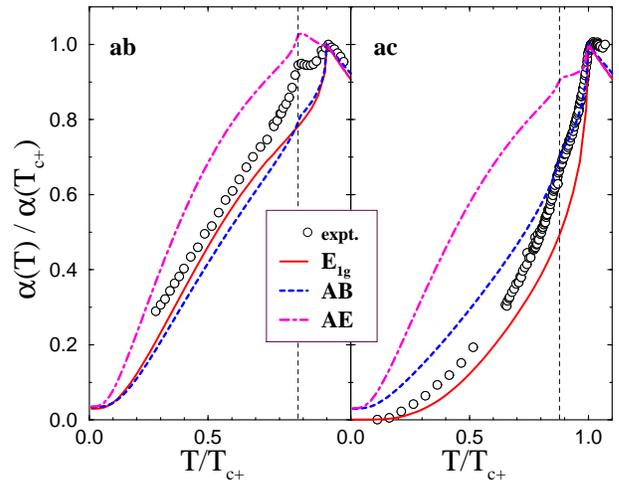}}}
\caption[]{Attenuation of $\va\vb$ and $\va\vc$ 
transverse sound for the various order parameter models.}
\label{theory}
\end{center}
\end{minipage}
\end{figure}

In Fig.~\ref{theory} we show the sound attenuation for
$\va\vb$ and $\va\vc$ polarizations and different pairing
models. Note in particular that none of the order parameter
models shown in Fig.~\ref{theory} can account
for both anomalies in $\alpha_{\va\vb}$ and $\alpha_{\va\vc}$ at $T_{c-}$.
The main result of this work is that the experimental data
for the heat capacity, thermal conductivity, and sound attenuation,
as well as the H-T phase diagram,
are explained {\sl only} by an order parameter with an orbital
E$_{2u}$ pairing symmetry.
We emphasize that there are no adjustable parameters
in the calculation of the sound attenuation; all parameters of the 
model were previously determined by fitting the theoretical 
model parameters
to the heat capacity and thermal conductivity.\cite{gra99}

\subsubsection{Domain Structure}

Neutron diffraction studies in pure UPt$_3$ under pressure,\cite{hay92}
and at ambient pressure in Pd doped samples 
(Pt $\leftrightarrow$ Pd),\cite{keizer99} demonstrated that the splitting of $T_c$
correlates with the basal plane AFM, suggesting a SBF coupling between the
superconducting and AFM order.
It has been argued, based on neutron diffraction studies as a function of 
magnetic field, that the AFM order is described either by a distribution of 
three equally populated domains, with $\bf Q$ vectors oriented $120^o$ to one 
another, or a triple-$\bf Q$ structure.\cite{lus96a,dij98} 
If a domain structure is present then, in the AFM model for the SBF,
the superconducting order parameter describing the A phase in the E-rep models 
may also form a domain structure. Such a domain structure leads to a weakening
of the anomaly for $\alpha_{\va\vb}$ at $T_{c-}$, but leaves the anomaly in
$\alpha_{\va\vc}(T_{c-})$ virtually unchanged.

\begin{figure}
\noindent
\begin{minipage}{87mm}
\begin{center}
\centerline{ \epsfxsize=0.75\hsize{\epsfbox{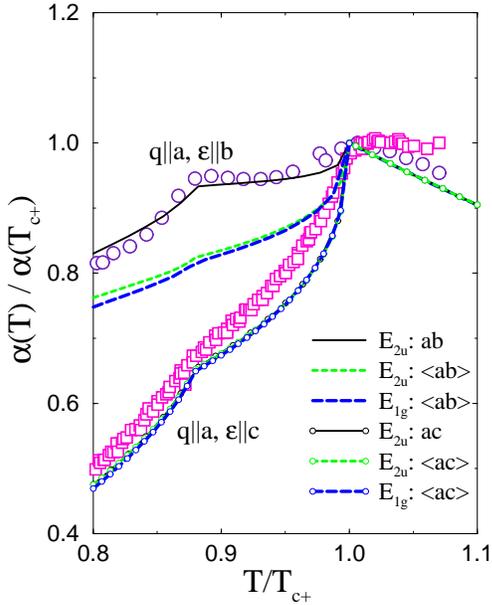}} }
\caption[]{Configuration averaged attenuation assuming three equally populated 
mono-domains for an E$_{1g}$ and E$_{2u}$ pairing state. For a comparison
the results for the E$_{2u}$ state of Fig.~\ref{theory} are re-plotted.
}
\label{multi-domain}
\end{center}
\end{minipage}
\end{figure}

In Fig.~\ref{multi-domain} we show the effect of multi-domain averaging 
on the anisotropy of the transverse sound attenuation.
These calculations show that an AFM domain structure does
not destroy the anomaly in the attenuation at $T_{c-}$.
However, averaging over domains suppresses the characteristic 
enhancement of $\alpha_{\va\vb}$ for the E$_{2u}$ order parameter
coupled to a dominant or single domain of
the SBF. In particular, the domain-averaged attenuation,
$\langle \alpha_{\va \vb} \rangle$, for the E$_{2u}$ pairing state 
drops roughly twice as fast as the measured attenuation in the A-phase.
If the AFM order parameter is the SBF for the superconducting phases,
then our calculations are in agreement with transport 
and heat capacity measurements only for an E$_{2u}$ order parameter
coupled to a {\sl dominant} domain, or a triple-$\bf Q$ structure 
for the antiferromagnetic SBF.\cite{mor00}

\subsubsection{Scattering phase shifts}

\begin{figure}
\noindent
\begin{minipage}{87mm}
\begin{center}
\epsfysize=0.90\hsize\rotate[r]{\epsfbox{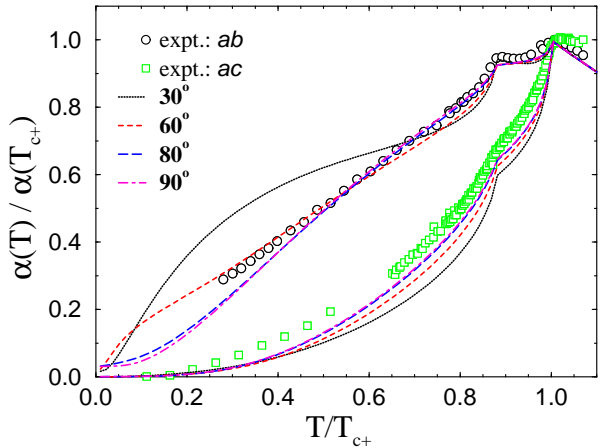}}
\caption[]{Sensitivity of the transverse sound attenuation to
the scattering phase shift 
$\delta_0 = 30^o, 60^o, 80^o, 90^o$ for the E$_{2u}$ pairing model
and the same parameters as shown in Fig. \ref{UPt3}.}
\label{Attenuation_phaseshift}
\end{center}
\end{minipage}
\end{figure}

It has been pointed out in several studies that the temperature dependence
of the transport coefficients in UPt$_3$ is qualitatively consistent with
strong scattering in the unitarity limit. Our analysis of the thermal
conductivity data,\cite{lus96b} and the attenuation data\cite{ell96} confirm
that the scattering phase shift is near the resonant
limit; from the analysis of the thermal conductivity we obtain
$\delta_0 \ge 80^o$,\cite{gra96b} 
while the transverse sound attenuation data implies
a scattering phase shift $\delta_0 > 60^o$, as shown in 
Fig.~\ref{Attenuation_phaseshift}.

\subsubsection{Ultra-low temperature region}

The temperature dependence of the transverse sound attenuation
is predicted to change
qualitatively below the cross-over temperature, $k_BT<\gamma$.
The zero-temperature limit is finite, reflecting the finite density 
of states at the Fermi level, and the leading temperature-dependent
corrections are  of the Sommerfeld type, ${\cal O}[(T/\gamma)^2]$
for $k_B T < \gamma$.
The limiting attenuation in the ultra-low temperature limit, $k_B T\ll\gamma$,
is obtained from the viscous stress tensor for $T\rightarrow 0$ and 
$\omega\rightarrow 0$. In this limit the
states contributing to the stress tensor in Eq.~(\ref{stress_tensor}) are
confined to the impurity-induced Andreev band of order
$\gamma \gg {\rm max}(T, \omega)$. The spectrum and
self-energy are weakly energy dependent on this scale, thus,
we can evaluate the slowly varying parts of the integrand in 
Eq.~(\ref{viscosity}) at zero energy to obtain
\begin{equation}\label{universal_viscosity}
\eta_{ij,ij} \simeq N_f\!
\int\!\!d\vp_f\,[{\bf v}_f]_i^2 [\vp_f]_j^2 \,
\frac{\gamma^2}{\left[|\Delta(\vp_f)|^2+\gamma^2\right]^{\tiny 3\over 2}}
\,,
\end{equation}
where we have neglected the Sommerfeld corrections of order
${\cal O}[(T/\gamma)^2]$.\cite{footnote1}

Whether or not the zero-temperature limit is {\sl universal}, i.e.,
independent of the density and scattering cross-section for the
impurities, depends on the polarization of the acoustic wave
and the symmetry of the ground state. 
The damping is universal only for transverse waves 
propagating along the high symmetry directions 
in the $ab$-plane, i.e., for polarization
\mbox{\boldmath$\varepsilon$}$ || \vb$ propagating along 
${\bf q} || \va$, or vice versa. For these polarizations the ground state
of {\sl all} the models we discuss, except the AE-model, possess a
{\it universal} limit for the attenuation. 
Table \ref{Limits} summarizes the results for the E$_{1g}$ and E$_{2u}$
models. For the ground state of either E-representation we find
$\eta_{\va\vb,\va\vb}\simeq v_f^2 p_f^2N_f/(8\mu\Delta_0)$,
where $\mu\approx 2$ is the slope of the excitation gap
near the line node in the $ab$-plane. For any other polarization the relevant 
viscosity is non-universal for $\omega, T\ll\gamma$. For example,
for the E$_{2u}$ ground state we obtain a limiting value for
$\alpha_{\va\vc}(0)\approx\frac{2\mu\gamma}{\mu_2^2\Delta_0}\alpha_{\va\vb}(0)
\ll\alpha_{\va\vb}(T_{c+})$,
where $\mu_2\approx 4$ is the parameter defining the curvature of the 
excitation gap near the quadratic point node along the 
$\vc$ axis.\cite{background}

Further experiments using transverse sound
can be used to confirm the predictions for the $ab$-plane
anisotropy of the order parameter in UPt$_3$.
The ideal method would be to propagate transverse sound along the
$\vc$ axis and measure the attenuation as a function of
the azimuthal orientation of the polarization in the $\va\vb$-plane.
The qualitative predictions for the in-plane anisotropy can
be deduced from Fig. \ref{NodeLocking-xy}; a two-fold symmetry
of the anomaly at $T_{c-}$ is expected for the E$_{1g}$ representation
and a nearly isotropic attenuation above $T_{c-}$ for the other models.
In the B phase all models show a nearly isotropic attenuation except for the
AE model, which has a two-fold symmetry.
A similar experiment was suggested by Moreno and Coleman\cite{mor96}
to map out the gap structure in the high-$T_c$ cuprates.

%
%
\noindent
\begin{minipage}{1.0\hsize}
\begin{table}
\caption{Asymptotic low-temperature limits of the
sound attenuation ($\omega$, $T\to 0$) and thermal conductivity
for the E$_{1g}$ and E$_{2u}$ pairing 
states. The results are scaled in units of
$\kappa_0=(\pi^2/3)k_B^2 T\,N_{\!f}v^2_{\!f}\tau_{\Delta}$ and
$\alpha_0=(\omega^2/4\varrho\,c_s^3)p^2_{\!f}\,N_{\!f}v^2_{\!f}\tau_{\Delta}$,
where $\tau_{\Delta}\equiv\hbar/2\mu\Delta_0(0)$ is
an effective transport time.}
\label{Limits}
\begin{tabular}{lll}
\noalign{\smallskip}
  {$\mbox{Transport}\atop\mbox{Coefficient}$}
  &     E$_{1g}$
  &     E$_{2u}$
\smallskip
\\ \hline
    $\kappa_{\vb\vb}/\kappa_{0}$
  & $\displaystyle {1}$
  & $\displaystyle {1}$
\\[1.0ex]
    $\kappa_{\vc\vc}/\kappa_{0}$
  & $\displaystyle {2\mu\gamma}/({\mu_1^2\Delta_0})$
  & $\!\displaystyle {\mu}/{\mu_2}\!$
\\[1.0ex]
    $\alpha_{\va\vb}/\alpha_{0}$
  & $\!\displaystyle {1}\!$
  & $\displaystyle {1}$
\\[1.0ex]
    $\alpha_{\va\vc}/\alpha_{0}$
  & $\displaystyle \frac{8\mu}{1+2\mu^2}\frac{\Gamma_u}{\Delta_0}^\star$
  & $\displaystyle \frac{2\mu}{\mu_2^2}\frac{\gamma}{\Delta_0}$
\\
\noalign{\smallskip}
\end{tabular}
{\footnotesize $^\star$ In the strong scattering limit 
               including vertex corrections.}
\end{table}
\end{minipage}

To summarize the main conclusions,
we have shown that transverse ultrasound provides detailed information 
on the orbital pairing symmetry of both the superconducting A and B phases
of UPt$_3$; the anisotropy and the anomalies in the temperature 
dependence of the attenuation for different polarizations
is explained {\sl only} by 
an E$_{2u}$ order parameter. Measurements of $\alpha_{\va\vb}$ and
$\alpha_{\va\vc}$ are in excellent agreement with a $(1,0)$ state 
for the A phase, corresponding to a $p_z (p_x^2 - p_y^2)$ 
order parameter.
Further measurements at lower temperatures, or as a function of 
impurity disorder, may also be used to test the prediction
of a universal limit for the in-plane transverse sound attenuation.

\section{Acknowlegdments}

  This research was supported by the NSF grant DMR 9705473 and
  DMR 91-20000 through the Science and Technology Center for
  Superconductivity. M.J.G. also acknowledges support from
  the Los Alamos National Laboratory under the auspices of the 
  Department of Energy.
  We also thank the Aspen Center for Physics where part
  of this research was carried out.

%
%

\appendix
\section{}

In this appendix we summarize the relevant nonequilibrium transport 
equations for our calculations of the sound attenuation in unconventional
superconductors. For a more extensive review of nonequilibrium 
transport theory in superconductors see Refs.~\onlinecite{gra96a,rai95}.

The nonequilibrium (Keldysh) Green's function,
\begin{equation}
\delta\hat{g}^K = \delta\hat{g}^R\circ\Phi_0-\Phi_0\circ\delta\hat{g}^A
                + \delta\hat{g}^a
\,,
\end{equation}
contains the {\sl spectral response} given by the retarded and
advanced Green's functions ($\delta\hat{g}^{R,A}$), and the
{\sl anomalous response} given by $\delta\hat{g}^a$, which 
in normal metals is essentially the nonequilibrium distribution
function. The equilibrium distribution function is 
$\Phi_0=\tanh({\epsilon}/{2 k_B T})$, and we use the shorthand 
notation for the shifting product,
$\Phi_0\circ A = \Phi_0(\epsilon-\omega/2) A(\epsilon,\omega)$ and
$A\circ \Phi_0 = A(\epsilon,\omega) \Phi_0(\epsilon+\omega/2)$.
Pairing correlations and particle-hole coherence require
a matrix structure for the
particle-hole degree of freedom. The `hat' over the Green's functions
and self-energies indicates their $4\times 4$ matrix 
structure in particle-hole and spin space.

The transport equations 
for the Green's functions,
linearized with respect to an external perturbation,
are
\begin{equation}
\left[\delta\hat{g}^{R,A},\,\hat{h}^{R,A} \right]_\circ = 
\left[\hat{g}_0^{R,A},\,\hat{\sigma}_{\rm ext}
     +\delta\hat{\sigma}^{R,A}\right]_\circ \,,
\end{equation}
\begin{eqnarray}
&&\hat h^R\circ\delta\hat{g}^{a}\,-\,\delta\hat{g}^{a}\circ\hat h^A =
\delta\hat{\sigma}^{a}\circ \hat g^A_0-\hat g^R_0\circ \delta\hat{\sigma}^{a}
\nonumber  \\
&&
-\left[ \hat{\sigma}_{\rm ext}\, , \,\Phi_0 \right]_\circ \circ\hat g^A_0
-\hat{g}^R_0\circ\left[\Phi_0\,,\,\hat{\sigma}_{\rm ext}\right]_\circ \,,
\end{eqnarray}
where $\hat{h}^{R,A}=\epsilon\hat{\tau}_3-\hat{\sigma}_0^{R,A}$,
$\hat{\tau}_3$ is a Pauli matrix, and $\hat{\sigma}_{\rm ext}$ is
the external perturbation, e.g., the coupling of quasiparticles
to a sound field (ionic displacement field).
The transport equations determine the deviations of the Green's
functions from their local equilibrium values,
$\delta\hat{g}^{X}=\hat{g}^{X}-\hat{g}^{X}_0$, in terms of the external
field and the corrections to the self-energies,
$\delta\hat{\sigma}^{X}=\hat{\sigma}^{X}-\hat{\sigma}^{X}_0$
with $X\in\{R,A,K\}$.
The anomalous self-energy, $\delta\hat{\sigma}^{a}$, is defined
similarly to the $\delta\hat{g}^{a}$,
\begin{equation}\label{sigmaK_lin}
\delta{\hat\sigma}^{K}=\delta\hat{\sigma}^{R}\circ\Phi_0
                      -\Phi_0\circ\delta\hat{\sigma}^{A}
                      +\delta\hat{\sigma}^a
\,.
\end{equation}
The equilibrium Green's functions, $\hat{g}_0^X$,
and self-energies, $\hat{\sigma}_0^X$, are inputs
to the linearized transport equations. At low temperatures, and for
long-wavelength, low-frequency sound, the damping of the
acoustic wave is determined by the scattering of quasiparticles
off impurities and defects that are co-moving with the
ionic lattice.\cite{tsu60a} Impurity scattering enters the
transport equations via the impurity-scattering self energies,
\begin{equation}
\hat{\sigma}^{X}(\vp_f;\epsilon,t)=n_{\mbox{\tiny
imp}}\,\hat{t}^{X}(\vp_f,\vp_f;\epsilon,t)
\,,
\end{equation}
where the quasiparticle-impurity scattering t-matrices are given by
\begin{eqnarray}
\hat{t}^{R,A}(\vp_f,\vp_f';\epsilon,t) &=&
	\hat{u}(\vp_f,\vp_f') + N_f \big\langle\hat{u}(\vp_f,\vp_f'')\circ
\nonumber \\ &&  \hspace{-8mm}
	\hat{g}^{R,A}(\vp_f'';\epsilon,t)\circ
	\hat{t}^{R,A}(\vp_f'',\vp_f';\epsilon,t)\big\rangle_{\vp_f''} 
\,,
\\
\hat{t}^{K}(\vp_f,\vp_f';\epsilon,t) &=&
	N_f\big\langle \hat{t}^{R}(\vp_f,\vp_f'';\epsilon,t)\circ
\nonumber \\ && \hspace{-8mm}
	\hat{g}^{K}(\vp_f'';\epsilon,t)\circ
	\hat{t}^{A}(\vp_f'',\vp_f';\epsilon,t)\big\rangle_{\vp_f''}
\,.
\end{eqnarray}
The t-matrices are calculated self-consistently with the order
parameter, $\hat{\Delta}(\vp_f)$, and with the impurity vertex,
$\hat{u}(\vp_f,\vp_f')$, describing elastic coupling of 
quasiparticles to impurities. The calculations presented here 
assume isotropic, non-magnetic impurities with $\hat{u}=u_0\hat{1}$
for all $(\vp_f,\vp_f')$. This model is then described by two parameters,
e.g., the s-wave scattering phase shift, $\delta_0=\tan^{-1}(\pi N_f u_0)$,
and the density of impurities, $n_{\mbox{\tiny imp}}$.
Another useful parametrization is in terms of the cross-section,
$\sigma=(4\pi/k_f^2)\sin^2\delta_0$, and the normal-state scattering
rate in the unitarity limit ($\delta_0=\pi/2$), $\Gamma_u=n_{\mbox{\tiny
imp}}/\pi N_f$. Note that the transport scattering rate in this
model is given by $1/2\tau_0=\Gamma_u\sin^2\delta_0$, and the elastic
mean free path is then $\ell_{\mbox{\tiny el}}=v_f\tau_0$.

The other key term entering the transport equations
is the off-diagonal pairing self energy, or order parameter.
The general form for the pairing self energies is
\begin{equation}
\hat{\Delta}^{R,A}=
\left(\matrix{ 0 & \Delta i\sigma_y + \vDelta\cdot i\vsigma\sigma_y \cr
\bar{\Delta} i\sigma_y + \bar{\vDelta}\cdot i\sigma_y\vsigma & 0}\right)\,,
\end{equation}
where the spin-singlet ($\Delta$) and spin-triplet ($\vDelta$) order parameters
are given by the gap equations,
\begin{eqnarray}
{\Delta}^{R,A}(\vp_f; t) &=& \int \frac{d\epsilon}{4 \pi i} 
\big\langle V^{s}(\vp_f,\vp_f')\,f^K(\vp_f';\epsilon,t)\big\rangle_{\vp_f'}
\,, \\
\vDelta^{R,A}(\vp_f;t) &=& \int \frac{d\epsilon}{4 \pi i} 
\big\langle \underline{\bf V}^{t}(\vp_f,\vp_f')\cdot {\bf f}^K(\vp_f';t)\big\rangle_{\vp_f'}
\,,
\end{eqnarray}
where $V^s$ and $\underline{\bf V}^t$ are the pairing interactions in the 
singlet and triplet channels.
The components $\bar{\Delta}$ and $\bar{\vDelta}$ are related
to $\Delta$ and $\vDelta$ by fundamental symmetries (see Appendix C of 
Ref.~\onlinecite{ser83}); 
in equilibrium $\bar{\Delta} = \Delta^*$ and $\bar{\vDelta} = \vDelta^*$.

To complete the set of equations for the linear
response equations we write the solutions for the equilibrium
response functions in terms of renormalized quasiparticle energy
and order parameter. For spin-singlet and unitary spin-triplet 
pairing the general solutions for the equilibrium retarded and advanced Green's 
functions are
\begin{equation}
\hat{g}_0^{R,A}=-\pi\frac{\tilde{\epsilon}^{R,A}\hat{\tau}_3
                 -\hat{\tilde{\Delta}}^{R,A}}
		 {\sqrt{|\tilde{\Delta}^{R,A}|^2-(\tilde{\epsilon}^{R,A})^2}}
\,,
\end{equation}
where the renormalized quasiparticle energy is
\begin{equation}
\tilde{\epsilon}^{R,A}(\vp_f,\epsilon)=\epsilon -
{\tiny 1\over 4}
\mbox{Tr}\left[\hat{\tau}_3\hat{\sigma}^{R,A}_{\mbox{\tiny imp}}(\vp_f,\epsilon)\right]
\,.
\end{equation}
For resonant s-wave scattering 
\begin{equation}
\hat{\sigma}^{R,A}_{\mbox{\tiny imp}}(\epsilon) =
-\Gamma_u \left<\hat{g}_0^{R,A}({\bf p}_f; \epsilon)\right>_{\vp_f}^{-1}
\,.
\end{equation}
For s-wave scattering the renormalization of the off-diagonal self-energy 
by impurity scattering vanishes,
\begin{equation}
\hat{\tilde{\Delta}}^{R,A}(\vp_f,\epsilon)=\hat{\Delta}(\vp_f)
\,,
\end{equation}
for non-identity representations of the point group, 
i.e., excluding
the A$_{1g}$ representation.
Finally, we note that the diagonal component of the nonequilibrium Green's
function that determines the momentum stress tensor in 
Eq.~(\ref{stress_tensor})
is obtained from the matrix Green's function by
$\delta g^{K}={\scriptsize 1\over 4}\mbox{Tr}\, \delta\hat{g}^{K}$.



\end{multicols}

\begin{thebibliography}{10}
\bibitem{coordinates}
We follow the notation for the crystallographic directions used
in experiments: $\va || \hat{x}$ and $\vb || \hat{y}$.
\bibitem{footnote2}
In the 2D E-rep models the splitting of $T_c$ is related to a weak symmetry breaking
field, either due to the antiferromagnetic order or structural modulations or
uniaxial stress.
\bibitem{background}
These asymptotic limits also imply that care must
be taken when analyzing the low-$T$ sound attenuation.
In particular a more sophisticated procedure for subtracting
the background attenuation is required than for conventional BCS
superconductors, where it is sufficient to demand that $\alpha$
vanishes as $T \to 0$.
\end{thebibliography}
\end{document}